\documentstyle[preprint,aps]{revtex}

\begin{document}

\tightenlines
\draft

\title{Topological dilaton black holes}

\author{Rong-Gen Cai } 
\address{Center for Theoretical Physics, Seoul National 
       University, Seoul 151-742, Korea}
\author{Jeong-Young Ji and Kwang-Sup Soh}
\address{Department of Physics Education, Seoul National University,
         Seoul 151-742, Korea}

\maketitle

\begin{abstract}
In  four-dimensional spacetime, when the two-sphere of black
hole event horizons is replaced by a two-dimensional hypersurface 
with zero or negative constant curvature, the black hole is 
referred to as a topological black hole. In  this paper we present 
some exact topological  black hole  solutions  in the 
Einstein-Maxwell-dilaton theory with a Liouville-type
dilaton potential.

\end{abstract}


The topological structure of the event horizon of a black hole 
is an intriguing subject in black hole physics.
It is generally believed that a black hole in the four dimensional 
spacetime is always with a spherical  topology.  That is, the event
horizon of a black hole has the topology $S^2$. This was proven 
by Friedman, Schleich and Witt \cite{FSW1}. They suggested the
``topological censorship theorem'', which states that in a globally
hyperbolic, asymptotically flat spacetime satisfying the null 
energy condition, every causal curve (nonspacelike curve) from 
${\cal J}^-$ to ${\cal J}^+$ is homotopic to a topologically trivial
curve from ${\cal J}^-$ to ${\cal J}^+$. That is, general relativity 
does not allow an observer to probe the topology of spacetime: Any 
topological structure collapses too quickly to allow light to traverse
it.   Later on, however, they found  that nontrivial  
topologies can be observed passively~\cite{FSW2}. 
The black holes with toroidal topology have 
indeed  been found numerically in the gravitational collapse
\cite{STW}, although such a topological structure is temporal.

When the asymptotic flatness  and the energy condition are given up,
there are no fundamental reasons
to forbid the existence of static or stationary black holes with
nontrivial topologies. In particular, when the spacetime is asymptotically 
anti-de Sitter one, the matter field can be in stable equilibrium even
if the potential energy is unbounded from below. 
In recent years,  there  has been a  growing
interest in these  black holes with nontrivial topological structures
(topological black holes) in the asymptotically anti-de Sitter 
space \cite{Lomes,Huang,Cai,Amin,Mann1,Smith,Bans,Mann2,Brill,Vanzo}. 
These investigations are mainly based on the 
Einstein (-Maxwell) theory  with a negative cosmological constant.
In general, one has the static solutions of  Einstein-Maxwell equations
with a cosmological constant
\begin{equation}
ds^2=-(k-\frac{2M}{r}+\frac{Q^2}{r^2} -\frac{1}{3}\Lambda r^2)dt^2
+ (k-\frac{2M}{r}+\frac{Q^2}{r^2} -\frac{1}{3}\Lambda r^2)^{-1}dr^2
+r^2d\Omega ^2_k,
\label{e1}
\end{equation}
where $d\Omega ^2_k$ is the line element of a two-dimensional
hypersurface $\Sigma $ with  constant curvature,
\begin{equation}
d\Omega ^2_k=\left \{
\begin{array}{ll}
d\theta ^2 +\sin^2\theta d\phi^2, & {\rm for}\ \ k=1 \\
d\theta ^2 + \theta ^2 d\phi ^2, & {\rm for}\ \ k=0 \\
d\theta ^2 +\sinh^2 \theta d\phi ^2, & {\rm for}\ \ k=-1.
\end{array} \right.
\label{e2}
\end{equation}
Here $M$ and $Q$ are the mass and charge of the solution, respectively,
and the cosmological constant $\Lambda$ is negative. 
For $k=1$, the metric (\ref{e1}) describes the spacetime of
the Reissner-Nordstr\"om-anti-de Sitter black holes.
Here the event horizon of the black hole has the 2-sphere
topology $S^2$, and the topology of spacetime is
$R^2\times S^2$. For $k=0$, if we identify the coordinates $\theta$ 
and $\phi $ with certain periods, the topology of event horizon 
is that of a torus and the spacetime has the topology $R^2 \times T^2$.
For $k=-1$, the surface $\Sigma $ is a 2-dimensional hypersurface with 
constant negative curvature. The topology of spacetime is $R^2\times H^2_g$
\cite{Smith}, where $H^2_g$ is the topology of the surface $\Sigma$.
Brill {\it et al.}\cite{Brill} have discussed in detail 
the topological structure of solution (\ref{e1}). 
In the case  of $k=-1$, the solution (\ref{e1}) is of some strange 
properties. Usually the occurrence  of black hole horizon is always 
related to the positive definiteness of energy of spacetime. 
In Eq. (\ref{e1}), however, even if $M=Q=0$, one has  the black hole
structure with  black hole horizon $r_h=\sqrt{3/|\Lambda}|$. In
particular, when the mass is negative, one still has the black hole 
solution. Surprisingly,  this negative mass black hole 
can also be  formed by
gravitational collapse \cite{Mann2}. Obviously, the asymptotically 
anti-de Sitter behavior 
 plays a crucial role in the existence of these nontrivial
topological black holes.

Note that only in Ref.~\cite{Cai} one of the present authors and
Zhang have considered briefly the dilaton black plane 
solutions ($k=0$).  In the present paper we would like to investigate 
the deformation of these topological black holes (\ref{e1}) 
by a dilaton field and a Liouville-type
dilaton potential (an effective cosmological constant). Due to the
dilaton field, the topological dilaton black holes found here will
become neither asymptotically anti-de Sitter nor asymptotically 
flat.

Consider the following action
\begin{equation}
S=\frac{1}{16\pi}\int d^4x\sqrt{-g}[R -2(\nabla \phi)^2 -
2\Lambda  e^{2b\phi}-e^{-2a\phi}F_{\mu\nu}F^{\mu\nu}],
\label{action}
\end{equation}
where $a$ and $b$ are  two constants  and $F_{\mu\nu}$ is the 
Maxwell field. We still  refer to $\Lambda$ as the
``cosmological constant''.
This action (\ref{action}) has been investigated 
in some detail by Chan {\it et al.}~\cite{Chan} in finding the 
spherically symmetric  black holes. So in this paper we will 
 consider the  cases  of $k=0$ and $k=-1$ only. Varying the action 
(\ref{action}) yields the equations of motion
\begin{eqnarray}
\label{eq4}
&& R_{\mu\nu}=2\partial _{\mu} \phi \partial _{\nu}\phi
     +g_{\mu\nu}\Lambda e^{2b\phi} +2 e^{-2a\phi}(F_{\mu \lambda}
	  F_{\nu}^{\ \lambda}-\frac{1}{4}g_{\mu\nu}F^2), \\
\label{eq5}	  
&&0=\partial _{\mu}(\sqrt{-g}e^{-2a\phi}F^{\mu\nu}), \\
\label{eq6}
&& \nabla ^2\phi= b\Lambda e^{2b\phi} -\frac{a}{2}e^{-2a\phi}F^2.
\end{eqnarray}
We assume the metric to be solved being of the form
\begin{equation}
ds^2 =-U(r)dt^2 +U^{-1}(r)dr^2 + R^2(r)d\Omega ^2_k.
\label{metric}
\end{equation}
Thus Eq. (\ref{eq5}) can be easily integrated to obtain
\begin{equation}
F_{tr}=\frac{4\pi Q}{V R^2}e^{2a\phi}.
\end{equation}
Here $V$ is the area of the hypersurface $\Sigma $ when it is
closed, and $Q$ is the 
electric charge defined as 
$Q=-\frac{1}{8\pi}\int e^{-2a\phi}\varepsilon
_{\mu\nu\alpha\beta}F^{\alpha\beta}$.
The equations (\ref{eq4}) and (\ref{eq6}) then can be 
simplified to
\begin{eqnarray}
\label{eq9}
&& -\frac{U''}{2}-\frac{R'U'}{R}= \Lambda
      e^{2b\phi}-\frac{16\pi ^2 Q^2}{V^2 R^4}e^{2a\phi},\\
\label{eq10}
&&-\frac{U''}{2}-\frac{R'U'}{R}-\frac{2R''U}{R}=2U\phi'^2 
       +\Lambda e^{2b\phi}-\frac{16\pi ^2 Q^2}{V^2 R^4}e^{2a\phi},\\
\label{eq11}		 
&&-\frac{1}{2R^2}\left [U(R^2)'\right ]'+\frac{k}{R^2}=\Lambda
      e^{2b\phi}+\frac{16\pi ^2 Q^2}{V^2 R^4}e^{2a\phi},\\
\label{eq12}		
&&\frac{1}{R^2}\left [R^2U\phi'\right ]'=\Lambda b e^{2b \phi}+
     \frac{16\pi ^2 aQ^2}{V^2 R^4}e^{2a\phi},
\end{eqnarray}	  
where a prime represents the derivative with respect to $r$. From Eqs.
(\ref{eq9}) and (\ref{eq10}) we have
\begin{equation}
\label{eq13}
R''/R=-\phi'^2.
\end{equation}
We further assume 
\begin{equation}
\label{eq14}
R(r)=\gamma r^N,
\end{equation}
where $\gamma$ and $N$ are  two constants. Such an assumption
(\ref{eq14}) has been extensively used to look for the dilaton
black hole solutions \cite{Chan,Chan1}. Thus from (\ref{eq13})
one has
\begin{equation}
\label{eq15}
\phi (r)=\phi _0 \pm \sqrt{N(1-N)}\ln r,  
\end{equation}
where $\phi _0$ is an integration constant. We now  discuss the 
cases of $k=0$ and $k=-1$, respectively.

(I) $k=0$: In this case we find two sets of solutions of physical 
interest. The first set is 
\begin{eqnarray}
\label{eq16}
&& U(r)=-\frac{8\pi M}{VN\gamma ^2 }r^{1-2N}-
      \frac{\Lambda e^{2b\phi _0}}{N(4N-1)}r^{2N}+
		\frac{16\pi ^2 Q^2e^{2a\phi_0}}{NV^2\gamma^4}r^{-2N},\\
\label{eq17}
&&\phi (r)=\phi _0 +\sqrt{N(1-N)}\ln r, \\		
\label{eq18}
&&a=b=\sqrt{N(1-N)}/N,
\end{eqnarray}
where $M$ is the quasilocal mass defined as in Refs.~\cite{Brown,Chan1}. 
 If  $a=b=-\sqrt{N(1-N)}/N$, the solution (\ref{eq16}) remains
 unchanged, but $\phi (r)$ becomes
$\phi (r)=\phi _0-\sqrt{N(1-N)}\ln r $.
From (\ref{eq18}) we must have $0<N<1$ (as $N=1$ one has only 
the trivial 
constant dilaton solution), and $N\ne 1/4$ [see Eq. (\ref{eq16})].

(i) $0 < N < 1/4$: The first term ($r^{1-2N}$) in (\ref{eq16})
is dominant as $r$ is very large. That is, $U(r)$ is negative as $r$ is
enough large. In this case, the solution 
(\ref{eq16}) will have a cosmological horizon, despite  the sign 
of the cosmological constant $\Lambda$. The horizons  
are determined by 
\begin{equation}
\label{eq19}
-\frac{8\pi M}{V\gamma ^2 }r -
      \frac{\Lambda e^{2b\phi _0}}{(4N-1)}r^{4N}+
		\frac{16\pi ^2 Q^2e^{2a\phi_0}}{V^2\gamma^4}=0,
\end{equation}
and due to $4N<1$, this equation has only a positive real root, 
which corresponds to the cosmological horizon.
The Hawking temperature of the horizon is 
\begin{equation}
\label{eq20}
T_h= \left |-\frac{2M}{VN\gamma ^2}r_c^{-2N}- 
   \frac{\Lambda e^{2b\phi _0}}{(4N-1)\pi}r_c^{2N-1}\right |,
\end{equation}
where $r_{\rm c}$ is the cosmological horizon. 

(ii) $1/4 <N<1$: The second term ($r^{2N}$) in the solution
(\ref{eq16}) will be dominant as $r\rightarrow \infty$. Due to 
$4N>1$, equation (\ref{eq19}) has at most two positive roots.
For $\Lambda >0$, equation (\ref{eq19}) clearly has only one solution, 
which corresponds to the cosmological horizon.
For $\Lambda <0$, the Eq.~(\ref{eq19}) has no cosmological horizon. 
But it may have the black hole horizons which are still determined by
the equation (\ref{eq19}). 
For example, for $N=1/2$, we have black hole horizons
\begin{equation}
r_{\pm}=\frac{4\pi M}{V\gamma^2 |\Lambda|e^{2\phi _0}}
      \left[1\pm \sqrt{1-\frac{Q^2|\Lambda| e^{4\phi _0}}{M^2}}
		 \right ].
\end{equation}		 
For $M^2 > |\Lambda |e^{4\phi _0}Q^2$,  
the solution (\ref{eq16}) has two black hole horizons, outer 
horizon $r_+$ and inner horizon $r_-$.
For $M^2=|\Lambda| Q^2e^{4\phi_0}$, the two horizons
coincide. This corresponds to the extremal topological
dilaton black hole. For $M^2 < |\Lambda |Q^2e^{4\phi _0}$, the
singularity at $r=0$ will be naked. Analytically continuing the black
hole solution to its Euclidean section, it is easy to find the Hawking 
temperature of the hole by requiring the absence of conical singularity
at the black hole horizon. 
The Hawking temperature of the black hole is found to be
\begin{equation}
T_h=\frac{|\Lambda|e^{2\phi _0}}{2\pi}-\frac{8\pi Q^2e^{2\phi_0}}
     {V^2\gamma^4r_+^2}.
\end{equation}
It is worth noting that when $Q=0$, the Hawking temperature will 
become a constant independent of the mass. This property is
very similar to that of the 2-dimensional charged dilaton black holes
\cite{Frolov}. For $N=3/4$, the solution (\ref{eq16}) may also have 
two black hole horizons. We can  obtain similar 
expressions of black hole 
horizons and the Hawking temperature as well.

The second set  of solutions is 
\begin{eqnarray}
\label{eq23}
&& U(r)=-\frac{8\pi M}{VN\gamma ^2 }r^{1-2N}
     +\frac{ 32\pi ^2 NQ^2 e^{2a\phi _0}}
	 {V^2\gamma^4 x(x+bN)(2bx+2N+1)}r^{2+2bx},\\
&&\phi (r)=\phi _0 +x\ln r,\\
&& (a-b)x=2N,\\
&& \Lambda =-\frac{x+aN}{x+bN}\frac{16\pi ^2 Q^2e^{2(a-b)\phi _0}}
    {V^2\gamma ^4},
\end{eqnarray}
where $x= \sqrt{N(1-N)}$. The solution (\ref{eq23}) has three
possibilities: naked singularity,  a cosmological horizon
 or a black hole horizon, depending on the parameters
$a$ and $b$. When the horizon is cosmological one, the singularity at
$r=0$ is a cosmological one.  The black hole horizon is
\begin{equation}
r_+=\left[\frac{4\pi N^2Q^2e^{2a\phi_0}}
        {V \gamma^2 x(x+bN)(2bx+2N+1)M}\right]^{-1/(1+2N+2bx)},
\end{equation}
and the associated Hawking temperature is
\begin{equation}
T_h=\frac{2(2N-1)M}{VN\gamma^2r^{2N}_+}
    +\frac{16\pi N(1+bx)NQ^2e^{2a\phi_0}r_+^{1+2bx}}
      {V^2\gamma^4 x(x+bN)(2bx+2N+1)}.
\end{equation}

(II) $k=-1$: In this case, we also find two sets of solutions of 
physical interest. The first is
\begin{eqnarray}
\label{eq29}
&&U(r)=-\frac{8 \pi M}{VN \gamma ^2}r^{1-2N}-\frac{\Lambda e^{2b
     \phi _0}}{(1-N)}r^{2-2N}+\frac{16\pi ^2 Q^2e^{2a\phi _0}}
	  {NV^2\gamma ^4}r^{-2N}   \\
&&\phi (r)=\phi _0 -\sqrt{N(1-N)}\ln r, \\
&&b=a^{-1}=N/\sqrt{N(1-N)},\\
\label{eq32}
&&\Lambda =-\frac{1-N}{1-2N}\frac{e^{-2b\phi _0}}{\gamma ^2}.
\end{eqnarray}
If $b=a^{-1}=-N/\sqrt{N(1-N)}$, the solution 
(\ref{eq29}) remains unchanged, but $\phi (r)=\phi _0+\sqrt{N(1-N)}
\ln r$. When $\Lambda >0$ and $ 1/2<N<1$ [see Eq. (\ref{eq32})], 
the solution (\ref{eq29}) has only one cosmological horizon
\begin{equation}
r_{\rm c}=\frac{4\pi (1-N)M}{VN\Lambda  e^{2b\phi _0}}
           \left[ -1 +\sqrt{1+\frac{N\Lambda  Q^2 
			  e^{2(a+b)\phi_0}}{(1-N)M^2}
			  }\right].
\end{equation}
When $\Lambda <0$ and $0<N<1/2$, we have two  black hole horizons
\begin{equation}
\label{eq34}
r_{\pm}=\frac{4\pi (1-N)M}{VN|\Lambda| \gamma^2 e^{2b\phi _0}}
           \left[ 1 \pm \sqrt{1-\frac{N|\Lambda| Q^2 
			  e^{2(a+b)\phi_0}}{(1-N)M^2}
			  }\right],  
\end{equation}
and the Hawking temperature is
\begin{equation}
T_h=\frac{2(2N-1) M}{VN\gamma^2 r^{2N}_+}
    +\frac{|\Lambda|e^{2b\phi_0}r^{1-2N}_+}{2\pi }
	 -\frac{8\pi Q^2e^{2a\phi_0}}{V^2\gamma^4 r_+^{2N+1}}.
\end{equation}

The second set of solutions is 
\begin{eqnarray}
\label{eq36}
&& U(r)=-\frac{8 \pi Mr^{1-2N}}{VN\gamma ^2}-\frac{\Lambda e^{2b\phi _0}
     r^{2-2N} }{(1-N)}+\frac{16\pi ^2 Q^2e^{2a\phi _0}r^{2-2N}}
	    {V^2 \gamma ^4(1-N)},\\
&&\phi (r)=\phi _0+\sqrt{N(1-N)},\\
&&a=-b=N/\sqrt{N(1-N)},\\
\label{eq39}
&&\Lambda =-\frac{(1-N)e^{-2b\phi _0}}{\gamma ^2(1-2N)}\left[
           \frac{16\pi ^2 Q^2e^{2a\phi _0}}{V^2 
			  \gamma ^2(1-N)}+1\right],
\end{eqnarray}
When $a=-b=-N/\sqrt{N(1-N)}$, the solution $U(r)$ keeps unchanged, but
$\phi (r)=\phi _0-\sqrt{N(1-N)}\ln r$.  When $\Lambda <0$, namely,
$0<N<1/2$ [see Eq. (\ref{eq39})], obviously, the solution (\ref{eq36})
has a black hole horizon
\begin{equation}
\label{eq40}
r_+=\frac{8\pi M}{VN\gamma ^2}\left[-\frac{\Lambda e^{2b\phi _0}}
      {(1-N)}+\frac{16\pi ^2 Q^2 e^{2a\phi _0}}{V^2\gamma ^4 (1-N)}
		\right]^{-1}.
\end{equation}
 The Hawking temperature is
\begin{equation}
T_h=\frac{2(2N-1)M}{VN\gamma^2r_+^{2N}}-\frac{\Lambda e^{2b\phi_0}
     r_+^{1-2N}}{2\pi} +\frac{8\pi Q^2e^{2a\phi _0}r_+^{1-2N}}
	  {V^2\gamma^4}.
\end{equation}

These topological dilaton  black hole solutions are 
 counterparts of  the spherically symmetric  black holes ($k=1$)
 \cite{Chan}.
 They are neither asymptotically flat nor (anti-)de Sitter, but 
 they have finite quasilocal masses and finite conserved charge 
 for compact event horizons.  These solutions have only a singularity 
 at $r=0$, and is enclosed
 by black hole horizons or cosmological horizon. In addition, 
 although these 
 topological dilaton black holes exhibit  unusual asymptotic 
 behavior, it is easy to show that  the entropy of black holes still 
 obeys  the area formula:
 \begin{equation}
 S=\frac{1}{4}VR^2(r_+)=\frac{1}{4}A(r_+),   
 \end{equation}
where $A(r_+)=VR^2(r_+)$ is the horizon area of topological black holes. 
This is because the black hole entropy comes from the surface term of the
Euclidean counterpart of action (\ref{action}) and the entropy is just
the value of the surface term at the black hole horizon \cite{Cai2}. 
 In the above we only analyze the 
causal  structure of
solutions for the positive quasilocal mass. If a negative 
quasilocal mass is allowed, there are richer black hole structures
for both cases of $k=0$ and $k=1$.
 Similar to that in Ref.~\cite{Chan}, the  generalization 
of the above topological dilaton black holes
to two Liouville-type potential $V=2\Lambda _1e^{2b_1\phi} 
+2\Lambda _2e^{2b_2\phi}$ ($b_1\ne b_2$) is a straightforward 
work, but the causal structure of solutions 
will become complicated. 
Due to the dilaton black hole solutions obtained  
by  Chan {\it et al.}~\cite{Chan}, our black hole
solutions presented here can be regarded as complements of those 
spherically symmetric  dilaton black holes. Finally,
it is worth noting that when $\Lambda =0$, the action (\ref{action})
allows the existence of the spherically symmetric black hole 
solutions, but does not for the topological dilaton black holes.

In summary, we have obtained some topological black hole solutions in 
the dilaton gravity with a Liouville-type dilaton potential. Their 
event horizons are the hypersurfaces with zero curvature or negative 
constant curvature. From these solutions we have found that the
effective cosmological constant in the action (\ref{action}) must still be
negative in order to have these topological black hole solutions.
This situation is quite different from that of the spherically symmetric 
dilaton black holes \cite{Chan}.

This work was supported by the Center for Theoretical Physics 
of Seoul National University.

\end{document}